\documentclass[reprint,superscriptaddress,amsmath,amssymb,floatfix]{revtex4-2}
\usepackage{newtxtext,newtxmath}
\usepackage{xcolor}
\usepackage{bm}
\usepackage{braket}
\usepackage{graphicx}
\definecolor{mygreen}{rgb}{0.0,0.65,0.0}
\definecolor{ascol}{rgb}{0.34,0.35,0.83}
\usepackage[pdfstartview=FitH, colorlinks=true, linkcolor=blue, citecolor=red, urlcolor=mygreen, pdfborder={0 0 0}]{hyperref}

\def\lmbd{\lambda}
\newcommand{\msc}[1]{\normalfont\textsc{#1}}
\def\ad{\msc{ad}}
\def\na{\msc{na}}
\def\th{\mathrm{th}}
\def\diss{\mathrm{diss}}
\def\hxc{\msc{Hxc}}
\def\bchi{\boldsymbol{\chi}}

\begin{document}

\title{Full Quantum Work Statistics for Non-Homogeneous Many-Body Systems}

\author{Antonio Palamara}
\email{corresponding author: antonio.palamara@unical.it}
\affiliation{Dipartimento di Fisica, Universit\`{a} della Calabria, Via P. Bucci, Cubo 30C, I-87036 Rende (CS), Italy}
\affiliation{INFN, Gruppo Collegato di Cosenza, Via P. Bucci, Cubo 31C, I-87036 Rende (CS), Italy}
\author{Francesco Plastina}
\affiliation{Dipartimento di Fisica, Universit\`{a} della Calabria, Via P. Bucci, Cubo 30C, I-87036 Rende (CS), Italy}
\affiliation{INFN, Gruppo Collegato di Cosenza, Via P. Bucci, Cubo 31C, I-87036 Rende (CS), Italy}
\author{Antonello Sindona}
\affiliation{Dipartimento di Fisica, Universit\`{a} della Calabria, Via P. Bucci, Cubo 30C, I-87036 Rende (CS), Italy}
\affiliation{INFN, Gruppo Collegato di Cosenza, Via P. Bucci, Cubo 31C, I-87036 Rende (CS), Italy}
\author{Irene D'Amico}
\affiliation{School of Physics, Engineering and Technology, University of York, YO10~5DD York, United Kingdom}
\affiliation{York Center for Quantum Technologies,  The University of York, YO10 5DD, York, United Kingdom}
\date{\today}

\begin{abstract}

The nonequilibrium thermodynamics of interacting quantum many-body systems is investigated within the framework of thermal time-dependent density functional theory using a generalized linear-response formulation for the full quantum work statistics.
A first-principles route is established to reconstruct the relaxation function that underlies linear-response theory, thereby moving beyond phenomenological descriptions and enabling a consistent evaluation of all moments of the dissipated-work distribution in interacting systems.
The predictive power of the approach is demonstrated for the Hubbard model subject to a staggered external potential, where the evolution of the relaxation dynamics during the Mott-to-band-insulator crossover reveals how distinct many-body phases shape the out-of-equilibrium thermodynamic response.
These results provide a microscopic and transferable framework for quantum thermodynamics in correlated systems, bridging thermal density functional theory and nonequilibrium work statistics.
\end{abstract}

\maketitle

\section{Introduction\label{sec:intro}}
Nonequilibrium thermodynamics in quantum systems has become a central theme in modern quantum statistical mechanics and quantum information science~\cite{Talkner2016, Campisi2011, Pekola2015, Vinjanampathy2016, Strasberg2017, Esposito2009RMP, campbell2025roadmap}, driven by fundamental questions concerning irreversibility and fluctuations at the quantum scale, as well as by the demands of emerging quantum technologies~\cite{Myers2022, Campaioli2024}.
In particular, the assessment of quantum fluctuations, which persist even in the absence of thermal noise, is essential for determining the energetic cost, performance, and stability of quantum protocols~\cite{Cleri2024, Preskill2018, Auffeves2022, Jaschke2023, Perciavalle2025, BlokLandi2025,PhysRevResearch.5.043104,PhysRevLett.127.030602}.
Over the past decade, ultracold atomic gases and trapped ions have emerged as highly controllable platforms for quantum simulations of condensed-matter phenomena~\cite{Bloch2012, Lewenstein2007, Bloch2008}.
Exploiting their intrinsic cleanliness and tunability, these systems have enabled precise experimental tests of quantum thermodynamic principles, including fluctuation relations and work statistics~\cite{Koch2023, Aimet2025, Onishchenko2024, Hu2020}.
These developments underscore the need for microscopic approaches to quantum thermodynamics in interacting many-body systems, where correlations and quantum coherence play a key role in shaping both equilibrium and nonequilibrium behavior~\cite{Geier2022, Hahn2023, Zhang2024, e28030297,fox2024harnessing, Jussiau2023, Jaramillo2016,PhysRevResearch.5.043104}.
Density functional theory~(DFT)~\cite{Hohenberg1964, Kohn1965} and its time-dependent extension~(TDDFT)~\cite{RungeGross1984, UllrichBook2012} provide a unified first-principles framework for describing interacting many-particle systems at zero temperature.
Originally formulated for the electronic many-body problem, these approaches have been generalized to lattice Hamiltonians and model systems~\cite{Schoenhhammer1995, Wu2006, Verdozzi2008, LiUllrich2008, Tokatly2011, Farzanehpour2012, Coe2015, Franca2018, Penz2021, Xu2022, CAPELLE201391}.
More recently, renewed attention has been devoted to thermal density functional theory~(thDFT)~\cite{Mermin1965, BSGP16, SPB16, YTPB14, PPFS11, PhysRevB.105.235114, PhysRevB.94.241103, Karasiev2016, Trickey2014} and its time-dependent extension~(thTDDFT)~\cite{PribramJones2016, van1999mapping}, motivated by the need to address finite-temperature properties and nonequilibrium processes in which thermal effects are indispensable.
In this setting, several works have explored density-functional formulations of quantum thermodynamics, aiming to connect nonequilibrium energy statistics with ground-state and time-dependent density functionals~\cite{Skelt2019, Zawadzki2022, Herrera2018, Herrera2017}.
Notably, a thermal density-functional framework for describing work distributions in sudden-quench processes was introduced~\cite{Palamara2024}, providing a first-principles route to express nonequilibrium thermodynamic observables directly in terms of thermal densities and Kohn-Sham potentials.

A central open question concerns the nature and energetic impact of quantum fluctuations during finite-time thermodynamic processes.
While quasi-static transformations have been extensively analyzed~\cite{PhysRevLett.126.210603, PhysRevLett.119.050601, PhysRevLett.124.110606, PhysRevB.102.155407, PhysRevLett.124.040602, PhysRevE.105.L052102, PRXQuantum.3.010326, Scandi2019thermodynamiclength, PhysRevLett.123.230603, PhysRevA.86.032114, PhysRevResearch.2.023377, onishchenko2024probing}, extending nonequilibrium formulations beyond this limit remains highly nontrivial, particularly for interacting many-body systems~\cite{PhysRevResearch.2.033167, PhysRevA.107.012209}.
A significant development introduced a generalized linear-response theory~(GLRT) for the full quantum work statistics~\cite{Guarnieri2024}, in which all moments of the dissipated-work distribution are expressed in terms of the relaxation function~\cite{Kubo1957, bonancca2014optimal}, encoding the response of the system to an external driving protocol.
When microscopic information is unavailable, the relaxation function may be modeled phenomenologically, underscoring the versatility of the framework, albeit at the expense of predictive accuracy.

In this work, a further advancement of this frontier is presented by deriving the relaxation function directly from microscopic first principles, thereby providing a predictive and transferable approach to quantum thermodynamic processes in interacting many-body systems.
Specifically, linear-response thermal time-dependent density functional theory~(LR-thTDDFT) is employed to obtain accurate approximations of the relaxation function and, consequently, of the full work distribution.
The applicability of the approach is illustrated for the Hubbard model in a staggered external potential by analyzing the first moments of the dissipated-work distribution across the Mott-to-band-insulator crossover.

In principle, thDFT and thTDDFT are formally exact; however, in practice, both equilibrium and nonequilibrium calculations rely on approximate exchange-correlation functionals~\cite{burke2013dft, burke2005time}.
Here, using the thermal adiabatic local density approximation~(thALDA)~\cite{PribramJones2016}, it is shown that even a local and temporally adiabatic kernel captures the essential role of many-body phases in shaping the nonequilibrium thermodynamics of the interacting system.

\section{Full quantum work statistics in linear response\label{sec:GLRT}}
To set the stage, consider a generic quantum system driven by a time-dependent external potential through the Hamiltonian
\begin{equation}
  \hat{H}(t) = \hat{H}_0 + \lmbd(t)\,\hat{V},
  \label{eq:Ht}
\end{equation}
where $\lmbd(t)$ varies over the interval $[0,\tau]$ and defines the driving protocol.
The operator $\hat{V}$ acts as a perturbation that is effectively switched on at $t=0$ and switched off at $t=\tau$.
Accordingly, the work parameter changes from $\lmbd(0)=0$ to $\lmbd(\tau)=\delta\lmbd$, with an average rate proportional to $1/\tau$.
The system is initially prepared in the thermal equilibrium state of $\hat{H}_0$ at inverse temperature $\beta$ and is subsequently driven unitarily into a nonequilibrium state by the protocol $\lmbd(t)$.

We focus on the dissipated-work distribution $P(W_{\diss})$, which underpins the full stochastic thermodynamics of the process, and in particular on its first moment $\langle W_{\diss} \rangle$, identified within the two-point measurement framework~\cite{Esposito2009RMP,TalknerLutzHanggi2007}. 
The dissipated work is strictly related to the irreversibility of the process and can be expressed exactly in terms of the quantum relative-entropy functional~$S(\cdot\,\Vert\,\cdot)$~\cite{Audenaert2014}, as
\begin{equation}
  \beta \langle W_{\diss} \rangle
  = S\big( \hat\rho_\tau \,\Vert\, \hat\rho_\tau^{\th} \big),
  \label{eq:Wdiss-relent}
\end{equation}
where $\hat\rho_\tau$ denotes the nonequilibrium state after the evolution, and $\hat\rho_\tau^{\th}$ is the thermal state toward which the system would relax at the end of the protocol~\cite{Deffner2011PRL}.

A central result of the GLRT~\cite{Guarnieri2024} is that the full dynamical response of the system is encapsulated in the \textit{relaxation function},
\begin{equation}
  \psi(t)
  = \beta \int_{0}^{\beta} ds\,
      \big\langle \hat{V}(-is)\,\hat{V}(t) \big\rangle_{0}
    - \beta^{2} \langle \hat{V} \rangle_{0}^{2},
  \label{eq:relaxation-function}
\end{equation}
where $\hat{V}(t)=e^{i\hat{H}_{0}t}\hat{V}e^{-i\hat{H}_{0}t}$ and $\langle \cdots \rangle_{0}$ denotes the thermal average with respect to $\hat{H}_{0}$.
This function satisfies two key constraints: time-reversal symmetry, $\psi(t)=\psi(-t)$, and the nonnegativity of its Fourier transform, $\tilde{\psi}(\omega)\ge0$, the latter ensuring consistency with the second law of thermodynamics~\cite{NazeBonanca2020}.

Furthermore, all cumulants $\{k^{n}_{\msc{w}}\}$ of the dissipated-work distribution are nonnegative and can be expressed as bilinear functionals of the driving protocol through the time derivative $\dot{\lambda}(t)$:
\begin{equation}
  \beta^{n} k^{n}_{\msc{w}}
  = \int_{-\infty}^{\infty}
    \frac{d\omega}{\sqrt{2\pi}}\,
    \tilde{\psi}(\omega)\,\gamma_{n}(\omega)\,
    \left|
      \int_{0}^{\tau} dt\,
      \dot{\lambda}(t)\,e^{i\omega t}
    \right|^{2}.
  \label{eq:kn-general}
\end{equation}
Here, $\tilde{\psi}(\omega)$ acts as a generalized friction kernel that encodes both the linear-response behavior and the thermodynamic irreversibility, while $\{\gamma_{n}\}$ are model-independent coefficients of the form
\begin{equation}
  \gamma_{n}(\omega)
  = \frac{1}{2}(\beta\omega)^{n-1}
  \begin{cases}
    \coth(\beta\omega/2), & \text{for even } n,\\[5pt]
    1, & \text{for odd } n.
  \end{cases}
  \label{eq:gamma}
\end{equation}
In particular, the irreversible entropy production follows from setting $n=1$ in Eqs.~(\ref{eq:kn-general}) and~(\ref{eq:gamma}), leading to the convolution integral
\begin{equation}
  \beta \langle W_{\diss} \rangle
  = \frac{1}{2}
    \int_{0}^{\tau} dt
    \int_{0}^{\tau} dt'\,
    \psi(t-t')\,\dot{\lambda}(t)\,\dot{\lambda}(t').
  \label{eq:Wdiss-1st}
\end{equation}

In the present work, we decompose each moment of the dissipated-work distribution into adiabatic and nonadiabatic contributions.
This separation is of fundamental importance, as it allows one to distinguish irreversibility originating from the time deformation of the energy spectrum (i.e., the adiabatic one), from that arising from genuinely dissipative nonadiabatic transitions.
Specifically, we show that the dissipation kernel in Eq.~(\ref{eq:Wdiss-1st}) can be expressed as a sum of adiabatic and nonadiabatic components,
\begin{equation}
  \psi(t) = \psi_{\na}(t) + \psi_{\ad},
  \label{eq:psi-split}
\end{equation}
where $\psi_{\na}$ contains all dynamical contributions arising from transitions between different energy levels and therefore governs genuine dissipative behavior.
The fundamental reason behind this statement lies in the fact that part of the irreversibility of a thermodynamic process arises from non-adiabatic internal friction~\cite{Kosloff2002,Plastina2014}. 
The remaining portion can be quantified by the irreversibility of a hypothetical adiabatic process.
The adiabatic contribution, $\psi_{\ad}$, is independent of the driving rate, involves no transition energies, i.e., no terms proportional to $\omega\ne0$, and reflects solely the difference between the instantaneous spectra of $\hat{H}(\tau)$ and $\hat{H}(0)$.

To obtain explicitly the decomposition of eq. (\ref{eq:psi-split}), let us consider a protocol in which the Hamiltonian evolves from $\hat{H}(0)$ to $\hat{H}(\tau)$ over an infinitely long time scale, $\tau\!\to\!\infty$, such that the conditions of the adiabatic theorem are satisfied~\cite{MessiahBook}.
Under these conditions, the adiabatic time-evolved density matrix takes the form
\begin{equation}
  \hat\rho_A
  = \sum_n p_n^0\,|n(\tau)\rangle\langle n(\tau)| ,
  \label{eq:rhoA}
\end{equation}
where $\{|n(\tau)\rangle\}$ denote the instantaneous eigenstates of the Hamiltonian at the end of the protocol~($t=\tau$), and $p_n^0 = e^{-\beta E_n(0)}/Z_0$ are the initial~($t=0$) equilibrium occupation probabilities for the eigenstates $|n(0)\rangle\equiv|n\rangle$, with $Z_0$ the partition function.

For an adiabatic process, using eq.(\ref{eq:Wdiss-relent}), one obtains
\begin{align}
  \beta \langle W_{\diss} \rangle_{\ad}
  &= S\big( \hat\rho_A \,\Vert\, \hat\rho_\tau^{\th} \big)\notag\\
  &= \sum_n p_n^0 \ln p_n^0
  + \beta \sum_n p_n^0 E_n(\tau) + \ln Z_\tau .
  \label{eq:Wdiss-ad-ent}
\end{align}
Assuming nondegeneracy of the instantaneous spectrum, and expanding the final energy levels perturbatively up to second order in $\delta\lmbd$ yields
\begin{equation}
  E_n(\tau)
  = E_n(0) + \delta\lmbd V_{nn}
    + \delta\lmbd^2
      \sum_{m\neq n}
      \frac{|V_{nm}|^2}
           {E_m(0)-E_n(0)} ,
  \label{eq:En-expansion}
\end{equation}
where $V_{nm} = \langle n|\hat{V}|m\rangle$, and, similarly, the logarithm of the partition function can be expanded as
\begin{align}
  \ln Z_\tau
  &= \ln Z_0
   - \beta \delta\lmbd \langle \hat V \rangle_0
   + \frac{\beta^2 \delta\lmbd^2}{2}
      \sum_n p_n^0 |V_{nn}|^2 \label{eq:lnZ-expansion}\\
  &\quad
   - \beta^2 \delta\lmbd^2
      \sum_n p_n^0
      \sum_{m\neq n}
      \frac{|V_{nm}|^2}{E_m(0)-E_n(0)}
   - \frac{\beta^2 \delta\lmbd^2}{2}
      \langle \hat V\rangle_0^2 .
\notag
\end{align}
Substituting Eqs.~\eqref{eq:En-expansion} and \eqref{eq:lnZ-expansion} into Eq.~\eqref{eq:Wdiss-ad-ent} gives the adiabatic contribution to the irreversible entropy production,
\begin{equation}
  \beta \langle W_{\diss} \rangle_{\ad}
  = \frac{\delta\lmbd^2\,\beta^2}{2}
    \Bigl(
      \sum_n p_n^0 |V_{nn}|^2 - \langle \hat V\rangle_0^2
    \Bigr).
  \label{eq:Wdiss-ad-final}
\end{equation}

Recalling Eq.~\eqref{eq:Wdiss-1st}, and noting that in the adiabatic limit the nonadiabatic part of the relaxation function must vanish identically due to the absence of transition processes, we obtain a time-independent expression for the adiabatic component,
\begin{equation}
  \psi_{\ad}
  = \beta^2
    \Bigl(
      \sum_n p_n^0 |V_{nn}|^2- \langle \hat V\rangle_0^2
    \Bigr).
  \label{eq:psi-ad}
\end{equation}
Consequently the nonadiabatic contribution reads
\begin{equation}
  \psi_{\na}(t)
  = \beta \int_0^\beta ds\,
     \big\langle \hat{V}(-is)\, \hat{V}(t) \big\rangle_0
    - \beta^2 \sum_n p_n^0 |V_{nn}|^2.
  \label{eq:psi-na}
\end{equation}

Finally, taking the Fourier transform of Eq.~\eqref{eq:psi-split} yields
\begin{equation}
  \tilde{\psi}(\omega)
  = \tilde{\psi}_{\na}(\omega)
  + \sqrt{2\pi}\,\delta(\omega)\,\psi_{\ad},
  \label{eq:psi_w_na_ad}
\end{equation}
so that, using Eq.~\eqref{eq:kn-general}, the adiabatic and nonadiabatic contributions to the moments of the dissipated-work distribution can be treated separately.
Owing to the structure of the coefficients $\{\gamma_n\}$, cf. Eq.~(\ref{eq:gamma}), the adiabatic term contributes only to the first two moments, consistent with the fact that, in the slow-quench limit, the dissipated-work distribution approaches a Gaussian form.

\section{relaxation function from LR-thTDDFT microscopic approach \label{sec:LRthTDDFT}}
Building upon the GLRT framework, we now demonstrate that LR-thTDDFT can, in principle, provide a first-principles approach to the relaxation function.
To keep the discussion general~\cite{Palamara2024}, consider a system described by the interacting Hamiltonian
\begin{equation}
  \hat{H}[\{\lambda_i\}]
  = \hat{H}_0 + \hat{V}_{\msc{ext}}[\{\lambda_i(t)\}]
  = \hat{H}_0 + \sum_{i=1}^{\mathcal{L}} \lambda_i(t)\,\hat{A}_i ,
  \label{eq:H-general}
\end{equation}
where $\hat{H}_0$ denotes that component of $\hat H$ which is independent of the external potential; i.e., in DFT terminology, the universal Hohenberg-Kohn~(HK) Hamiltonian~\cite{Hohenberg1964, Franca2018} and $\hat{V}_{\msc{ext}}$ represents a time-dependent external potential governed by a set of $\mathcal{L}$ dynamic control parameters $\lambda_i(t) = [\lambda_i^0 + \delta\lambda_i(t)]$. 
This general form of $\hat{H}$ encompasses a broad range of systems, including spin chains as well as fermionic and bosonic models.
The central quantity in TDDFT is the time-dependent expectation value
\begin{equation}
  a_i(t) := \mathrm{Tr}\!\left\{\hat{\rho}(t)\,\hat{A}_i\right\},
  \label{eq:ai-def}
\end{equation}
where $\hat{\rho}(t)$ is the unitarily evolved density operator originating from the initial grand-canonical or canonical thermal state $\hat{\rho}_{\beta}^{\th}[\{\lambda_i\}]$.
The Runge-Gross (RG) theorem~\cite{RungeGross1984} establishes, under fairly general conditions and with certain subtleties for lattice systems~\cite{LiUllrich2008, Farzanehpour2012, Coe2015}, a one-to-one correspondence between the set of time-dependent ``\textit{densities}'' $\{a_i(t)\}$ and the external controllers $\{\lambda_i(t)\}$, and thus with $\hat{V}_{\msc{ext}}[\{\lambda_i(t)\}]$, for a given initial state.
Furthermore, the van Leeuwen theorem~\cite{van1999mapping} guarantees the
existence of a non-interacting reference system evolving under an effective
potential that reproduces the same density dynamics as the interacting
system. This potential, referred to as the time-dependent Kohn-Sham (KS)
potential, incorporates both the external driving and the
Hartree-exchange-correlation~(Hxc) contributions.
In the lattice formulation, these contributions appear as
site-resolved KS fields, which should not be confused with the thermodynamic
work parameters, and are denoted by $\lambda_{i,\hxc}[\{a_i(t)\}]$.
Accordingly, the KS work parameters take the form
\begin{equation}
  \lambda_{i,\mathrm{KS}}[\{a_i(t)\}]
  = \lambda_i(t) + \lambda_{i,\hxc}[\{a_i(t)\}],
  \label{eq:LambdaKS}
\end{equation}
where $\lambda_{i,\mathrm{KS}}(t)$ is the discrete analogue of
the continuum KS potential, now acting on the
local operators $\hat A_i$.
As a consequence, the interacting Hamiltonian~(\ref{eq:H-general}) is replaced
by the KS Hamiltonian
\begin{equation}
  \hat{H}_{\mathrm{KS}}(t)
  = \hat{H}^{\mathrm{KS}}_{0}
  + \sum_{i=1}^{\mathcal{L}} \lambda_{i,\mathrm{KS}}(t)\,\hat{A}_i ,
  \label{eq:H-KS}
\end{equation}
where $\hat H_{0}^{\mathrm{KS}}$ plays the role of the one-body kinetic operator in the continuous TDDFT formulation and generates the non-interacting reference
dynamics.

For sufficiently weak time-dependent perturbations, the system enters the linear-response regime, and the changes in the densities can be obtained through the  density-density~(i.e., the $\hat{A}_j$-$\hat{A}_i$) response function, which  is defined in the time domain as
\begin{equation}
  \chi_{ij}^\beta(t)
  = -i\,\theta(t)\,
    \big\langle [\hat{A}_i(t), \hat{A}_j(0)] \big\rangle_0,
  \label{eq:chiAA}
\end{equation}
where $\theta(t)$ denotes the Heaviside step function.
Its frequency-domain form reads
\begin{equation}
  \chi_{ij}^\beta(\omega)
  = \sum_{n,m} p_n
  \left[
    \frac{\langle n|\hat{A}_i|m\rangle\langle m|\hat{A}_j|n\rangle}
         {\omega - \omega_{mn} + i0^{+}}
   - \frac{\langle n|\hat{A}_i|m\rangle\langle m|\hat{A}_j|n\rangle}
         {\omega + \omega_{mn} + i0^{+}}
  \right],
  \label{eq:chiAAw}
\end{equation}
which generalizes the Lehmann representation~\cite{Lehmann1954} to finite temperatures.
Here, $p_n$ is the thermal occupation probability of the $n$th eigenstate of the Hamiltonian~(\ref{eq:H-general}) and $\omega_{mn}=E_m-E_n$ are the corresponding energy differences.
The single-particle density~($\hat{A}_i$)-density~($\hat{A}_j$) response function for the non-interacting KS system, $\chi_{ij \msc{KS}}^\beta(\omega)$, can be defined in an analogous manner. The strength of the KS framework lies in the fact that, in the linear-response regime, the interacting response function can be obtained from the KS one through the Gross-Kohn (GK) formalism~\cite{PribramJones2016,GrossKohn1985}, leading to the frequency-domain response equation

\begin{equation}
  \boldsymbol{\chi}^\beta(\omega)
  = \boldsymbol{\chi}_{\msc{KS}}^\beta(\omega)
  + \boldsymbol{\chi}_{\msc{KS}}^\beta(\omega)\!\ast
    \mathbf{f}_{\hxc}^{\beta}(\omega)\!\ast
    \boldsymbol{\chi}^\beta(\omega),
  \label{eq:Gross-Kohn}
\end{equation}
where spatial indices have been omitted for clarity, bold symbols denote matrices in the site indices, and $\ast$ indicates matrix convolution.
Equation~\eqref{eq:Gross-Kohn} represents a formally exact Dyson-like relation that connects the interacting response function to its KS counterpart via the finite-temperature Hxc kernel $\mathbf{f}_{\hxc}^{\beta}(\omega)$~\cite{UllrichBook2012},
which is defined as
\begin{equation}
  \boldsymbol{f}^{\beta}_{\hxc}(\omega)
  =
  \bigl[\boldsymbol{\chi}_\beta(\omega)\bigr]^{-1}
  - \bigl[\boldsymbol{\chi}^{\mathrm{KS}}_\beta(\omega)\bigr]^{-1}.
  \label{eq:fxc-def}
\end{equation}
Following the Kubo-formula approach, the response function, in particular its non-adiabatic component, can be directly related to the relaxation function through its time derivative~\cite{Kubo1957,bonancca2014optimal,PhysRevE.92.042148}:
\begin{align}
  \frac{d\psi(t)}{dt}
  &= \frac{d\psi_{\na}(t)}{dt}
   = -i\beta \big\langle [\hat{V}(t),\hat{ V}(0)] \big\rangle_0
  \nonumber\\
  &= -i\beta \sum_{ij} g_ig_j\,
     \big\langle [\hat A_i(t),\hat A_j(0)] \big\rangle_0 .
  \label{eq:dpsi-dt}
\end{align}
where, in accordance to eq. (\ref{eq:Ht}), we have used $\lambda_i(t) = [\lambda_0 + \delta\lambda(t)]\,g_i$, with $g_i$ encoding the spatial pattern.
On the other hand, the same \textit{density-density} response function has been given in Eq.~(\ref{eq:chiAA}).
Comparing the Fourier transforms of Eqs.~\eqref{eq:dpsi-dt} and~\eqref{eq:chiAAw} yields a direct correspondence between $\tilde{\psi}_{\na}(\omega)$ and $\bchi_\beta(\omega)$,
\begin{equation}
  \tilde{\psi}_{\na}(\omega)
  = 2\beta \sum_{ij}g_ig_j
           \frac{\mathrm{Im}\,\chi^\beta_{ij}(\omega)}{\omega}.
  \label{eq:Psi-na-chi}
\end{equation}
Thus, the nonadiabatic component of the relaxation function is determined by the imaginary part of $\chi^\beta_{ij}(\omega)$. This could have been expected, as $\mathrm{Im}\,\chi^\beta_{ij}(\omega)$ directly reflects the transition probabilities mediated by the operators $\hat{A}_i$.
Upon comparing Eqs.~(\ref{eq:psi_w_na_ad}) and~(\ref{eq:Psi-na-chi}), the total relaxation function in the frequency domain for a nonhomogeneous many-body system can be expressed as
\begin{equation}
  \tilde{\psi}(\omega)
  = 2\beta \sum_{ij}g_ig_j
            \frac{\mathrm{Im}\,\chi^\beta_{ij}(\omega)}{\omega}
    + \sqrt{2\pi}\,\delta(\omega)\,\psi_{\ad},
  \label{eq:Psi-total}
\end{equation}where the $\delta(\omega)$ contribution arises from the time-independent
adiabatic component $\psi_{\ad}$ in Eq.~\eqref{eq:psi-ad}, which is independent of both time and quench rate, carries no transition energies and
therefore contributes only at zero frequency.
By virtue of the HK and KS theorems, the second term in Eq.~\eqref{eq:psi-ad}, being a sum of products of thermal densities, is, in principle, exact when evaluated within the KS system, whereas the first term must be approximated.
One possible strategy is to estimate the leading nontrivial contribution using many-body perturbation theory, taking the KS system as the reference.
Alternatively, a useful formal expression for $\psi_{\ad}$ can be obtained by exploiting the result for the irreversible entropy production in a sudden-quench protocol~\cite{Palamara2024}:
\begin{equation}
  \beta k^{1}_{\msc{w}}
  = -\frac{\beta^2}{2}
    \sum_{ij} \delta \lmbd_i\, \delta \lmbd_j\,
    \frac{\partial a_i^{\beta,0}}{\partial \lmbd^0_j},
  \label{eq:k1-sudden}
\end{equation}
where $a_i^{\beta,0}=\mathrm{Tr}\!\left\{\hat\rho_\beta^{\th}[\{\lambda_i^0\}]\,\hat A_i\right\}$ is the equilibrium thermal density.

Eq.~\eqref{eq:k1-sudden} coincides with the irreversible entropy production obtained from Eq.~\eqref{eq:kn-general} in the $\tau\!\to\!0$ limit for $n=1$.
Solving the resulting equality for $\psi_{\ad}$ yields
\begin{align}
  \psi_{\ad}
  =&
  -\frac{\beta^2 \delta\lmbd^2}{b(0)}
   \sum_{ij} g_i g_j\,
   \frac{\partial a_i^{\beta,0}}{\partial \lmbd_j}
  \label{eq:psi-ad-final}\\
  &- 2\beta
   \sum_{ij} g_i g_j
   \int \frac{d\omega}{\sqrt{2\pi}}\,
     \frac{\mathrm{Im}\,\chi^\beta_{ij}(\omega)}{\omega}\,
     \frac{b(\omega)}{b(0)},
\nonumber
\end{align}
where
\begin{equation}
  b(\omega)
  = \lim_{\tau \to 0}
    \left|
      \int_{0}^{\tau} dt\,
      \dot{\lmbd}(t)\,e^{i\omega t}
    \right|^2,
  \label{eq:bomega}
\end{equation}
represents the spectral weight of the driving protocol, quantifying how the sudden temporal profile of the control parameter distributes energy across frequencies and couples to the intrinsic excitation modes of the system. 
Importantly, $b(\omega)$ depends solely on the driving protocol and is independent of the system Hamiltonian.

Equation~(\ref{eq:Psi-total}) was previously derived in Ref.~\cite{vn83-mt2v}, where the work statistics of an interacting many-body system under finite-time driving were evaluated exactly within linear response. 
While the formal structure is identical, the present work provides a transparent physical interpretation of its components in terms of adiabatic and non-adiabatic contributions to the dissipated work. In particular, the term involving the imaginary part of $ \chi_{ij}^\beta$ yields the non-adiabatic contribution, which vanishes in the adiabatic limit, whereas the delta-function term captures the dissipated work associated with infinitely slow driving. The present formulation also naturally extends to multi-parameter driving protocols.

In summary, this section establishes that LR-thTDDFT provides the relaxation function through the following steps: 
(i) computation of the KS response function; 
(ii) inclusion of interaction effects via the Hxc kernel to obtain the fully interacting response; 
(iii) extraction of $\mathrm{Im}\,\chi_{ij}^\beta(\omega)$ through Eq.~(\ref{eq:Gross-Kohn}); 
and (iv) combination of the non-adiabatic contribution with the static adiabatic term to obtain the full $\tilde{\psi}(\omega)$.

It is worth emphasizing that the formalism developed up to this point is exact within the linear-response regime. 
In practice, however, approximations enter at two distinct levels. 
The first concerns the Hxc potential, determined by $\lambda_{i,\msc{KS}}[\{a_i(0)\}]$, which defines the self-consistent equilibrium KS Hamiltonian $\hat{H}_{\msc{KS}}[\{\lambda_{i,\msc{KS}}(0)\}]$ in Eq.~(\ref{eq:H-KS}) and the corresponding response function $\boldsymbol{\chi}_{\mathrm{KS}}^\beta(\omega)$. 
The second concerns the approximation adopted for the dynamical Hxc kernel defined in Eq.~(\ref{eq:fxc-def}). The most widely used choice is the thermal adiabatic local-density approximation (thALDA)~\cite{PribramJones2016}, in which the kernel is instantaneous in time, i.e., $f^{\beta}_{Hxc}(\omega)\approx f^{\beta}_{Hxc}$. 
This approximation is employed throughout the remainder of this work.

\section{Out-of-equilibrium quantum thermodynamics of the Hubbard model\label{sec:Hubbard}}

\begin{figure}
  \centering
  \includegraphics[width=0.65\columnwidth]{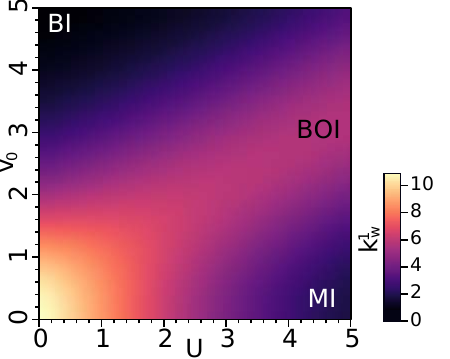}
\caption{Signature of the Hubbard-model phase diagram in the presence of a staggered field, as extracted from the dissipated work, $k^1_w=$$\langle W_{\diss}\rangle/(\delta v)^2$, following a sudden quench of the control parameter. Results are reported for inverse temperature $\beta J = 1$ and quench amplitude $\delta v = 0.01\,J$ in a half-filled chain of length $\mathcal{L} = 50$.}
  \label{fig:phase-diagram}
\end{figure}
The Hubbard model~\cite{EsslerBook} provides an ideal testbed for the LR-thTDDFT framework developed in the previous section. It can be experimentally realized in ultracold atomic gases using a superlattice potential~\cite{Pertot2014} and captures a wide variety of correlated-electron phenomena.

To keep the discussion focused, we consider the half-filled case with total spin projection along the $z$ axis equal to zero, subject to a time-dependent external potential.
The Hamiltonian for a finite chain of length $\mathcal{L}$ has the form
\begin{equation}
  \hat{H}(t) = \hat{T} + \hat{W} + \hat{V}_{\msc{ext}}(t),
  \label{eq:H-Hubbard}
\end{equation}
where
\begin{equation}
  \hat{T}
  = -J \sum_{i=1}^{\mathcal{L}} \sum_{\sigma=\uparrow,\downarrow}
    \left(
      \hat{c}_{i,\sigma}^\dagger \hat{c}_{i+1,\sigma}
      + \mathrm{h.c.}
    \right)
  \label{eq:T-Hubbard}
\end{equation}
is the kinetic hopping term,
\begin{equation}
  \hat{W}
  = U \sum_{i=1}^{\mathcal{L}} \hat{n}_{i,\uparrow}\hat{n}_{i,\downarrow}
  \label{eq:W-Hubbard}
\end{equation}
is the on-site Hubbard interaction, and

\begin{equation}
  \hat{V}_{\msc{ext}}(t)
  = \sum_{i=1}^{\mathcal{L}} v_i(t) \hat{n}_i,
  \quad  v_i(t) = v(t) (-1)^i,
  \label{eq:Vext-Hubbard}
\end{equation}
is a staggered, time-dependent external potential characterized by the work parameter $v(t) = v_0 + \frac{\delta v\,t}{\tau}$, which drives the system out of equilibrium over the interval $t\!\in\![0,\tau]$, with $\delta v_i = \delta v (-1)^i$ and $\delta v \!\ll\! v_0$.
In Eqs.~(\ref{eq:H-Hubbard})-(\ref{eq:Vext-Hubbard}), $\hat{c}_{i,\sigma}^\dagger$, $\hat{c}_{i,\sigma}$, and $\hat{n}_{i,\sigma}$ denote the creation, annihilation, and number operators for a fermion at site $i$ with spin $\sigma=\uparrow,\downarrow$, respectively, while $\hat{n}_i=\hat{n}_{i,\uparrow}+\hat{n}_{i,\downarrow}$ is the total number operator at site $i$.

The model Hamiltonian~(\ref{eq:H-Hubbard}) exhibits a rich ground-state phase diagram. In the $(U,v_0)$ plane, three insulating phases can be identified: a Mott-insulating (MI) phase, favoring single occupancy of lattice sites; a band-insulating (BI) phase, promoting double occupancy due to the external staggered potential; and a bond-order insulating (BOI) phase characterized by spontaneous dimerization. 
The latter emerges from the competition between the electron--electron interaction~(\ref{eq:W-Hubbard}) and the external staggered potential~(\ref{eq:Vext-Hubbard}), and is bounded by two critical lines. In particular, the BI--BOI transition is of Ising type, and the corresponding critical line follows approximately $U \sim 2v_0$ in the strong-coupling regime $U, v_0 \gg J$~\cite{PhysRevLett.83.2014,Tincani2009,PhysRevB.70.155115}.
In the following, we show that LR-thTDDFT provides direct access to this interplay, enabling the exploration of the system’s nonequilibrium quantum thermodynamics from the sudden-quench regime to the adiabatic limit, in the vicinity of a phase-transition precursor.
We specialize to the spin-independent density-density response function
$\chi_{ij}(t) = -i\,\theta(t)\langle[\hat{n}_i(t),\hat{n}_j]\rangle_0$,
as given in Eq.~(\ref{eq:chiAA}), and adopt the thALDA form
\begin{equation}
  f^{\beta\,ij}_{\hxc}[n_i^\beta]
  =
  \frac{\partial v^{\beta}_{\hxc}[n_i^\beta]}{\partial n_j^\beta}\,
  \delta_{ij},
  \label{eq:fxc-thALDA}
\end{equation}
for the Hxc kernel, derived from the Hxc potential
\begin{equation}
  v^{\beta}_{\hxc}[n_i^\beta]
  = U + \frac{1}{\beta}\ln\Gamma^\beta_U[n_i^\beta],
  \label{eq:vhxc-Hubbard}
\end{equation}
recently introduced to describe the quantum thermodynamics of the Hubbard model in the sudden-quench regime~\cite{Palamara2024}, where
\begin{equation}
  \Gamma^\beta_U[n_i^\beta]
  =
  \frac{
    (n_i^\beta-1) +
    \sqrt{(n_i^\beta-1)^2 + e^{-\beta U}\bigl[1-(n_i^\beta-1)^2\bigr]}
  }{n_i^\beta}.
  \label{eq:Gamma-Hubbard}
\end{equation}
Under this approximation, the exchange-correlation kernel becomes frequency independent.
As a consequence, the theory neglects memory effects and, crucially, cannot describe excitonic features beyond single particle hole-excitation that may arise in correlated many-body systems.
In other words, within this approximation, the number of poles in the interacting density-density response function coincides with that of the noninteracting Kohn-Sham system.
Though Eq.~(\ref{eq:fxc-thALDA}) neglects nonlocal correlations in both space and time, it still yields highly accurate results, in excellent agreement with exact diagonalization of the two-site Hubbard model, as shown in Appendix~\ref{app:dimer}.

The starting point for obtaining the relaxation function from the density-density response function within LR-thTDDFT is the self-consistent diagonalization of the KS Hamiltonian,
\begin{equation}
  \hat{H}_{\msc{KS}}
  =
  \hat{T}
  + \sum_i v_i^{\msc{KS}}[n_i^\beta]\,\hat{n}_i,
  \label{eq:KS-eq}
\end{equation}
which provides  the equilibrium thermal occupations of the initial Hubbard Hamiltonian $\hat{H}(0)$ in Eq.~(\ref{eq:H-Hubbard}). The KS system is constructed in the same ensemble as the interacting system.
The thermal site occupations are given by
\begin{equation}
  n_i^\beta
  = -\frac{2}{\beta}
    \frac{\partial \ln Z^{\msc{KS}}_{N}}{\partial v_i^{\msc{KS}}},
  \label{eq:thermal-density}
\end{equation}
where $v_i^{\msc{KS}}[n_i^\beta]$ denotes the KS potential and $Z^{\msc{KS}}_{N}$ is the grand-canonical partition function of the KS system with $N_\uparrow$ spin-up and $N_\downarrow$ spin-down fermions, such that $N = N_\uparrow + N_\downarrow$.
\begin{figure*}
  \centering
  \includegraphics[width=\textwidth]{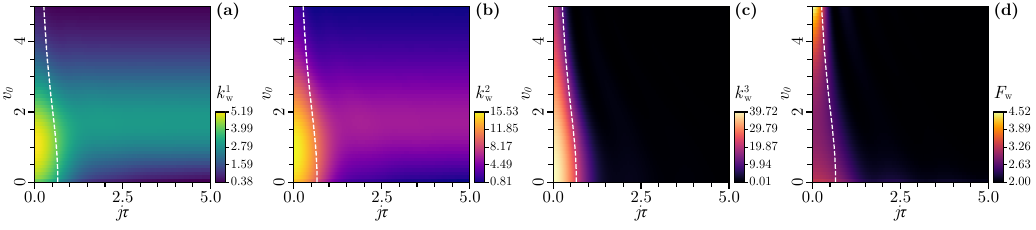}
  \caption{(a)-(c) First three cumulants of the dissipated-work distribution for the Hubbard Hamiltonian~(\ref{eq:H-Hubbard}) and (d) the corresponding Fano factor, plotted as functions of the quench duration $\tau$ and the staggered potential amplitude $v_0$. Results are shown for inverse temperature $\beta J = 1$, quench amplitude $\delta\lambda = 0.01 J$, $U=1J$ and a half-filled chain of length $L = 50$. The dashed line $\tau^\ast(v_0)$ indicates the crossover between the nonadiabatic and adiabatic regimes, identified as the maximum of the absolute derivative of $k^{3}_{\msc{w}}$ with respect to the quench time.}
  \label{fig:cumulants}
\end{figure*}
Consistent with the thALDA form in Eq.~(\ref{eq:fxc-thALDA}), we adopt a thermal local-density approximation~(thLDA) constructed from the Hxc functional of Eq.~\eqref{eq:vhxc-Hubbard},
\begin{equation}
  v_i^{\msc{KS}}[n_i^\beta]
  \simeq v_i + v^{\beta}_{\hxc}[n_i^\beta],
  \label{eq:vKS-Hubbard}
\end{equation}

On the simulation side, we considered chains of length $\mathcal{L}=50$ with an average particle number $N=50$.
As a first step, we analyzed the irreversibility generated by a sudden quench of the external potential, $v_i(0) \!\to\! v_i(0)+\delta v_i$, starting from a grand-canonical Gibbs state.
Figure~\ref{fig:phase-diagram} shows the behavior of the first moment of the dissipated work in the $(U,v_0)$ plane,
\begin{equation}
  \beta\,\frac{\langle W_{\diss}\rangle}{(\delta v)^2}
  =
  -\frac{\beta^2}{2}
   \sum_{ij}(-1)^{i+j}
   \frac{\partial n_i^\beta}{\partial v_j}, 
\end{equation}
as obtained from Eq.~(\ref{eq:k1-sudden}), which expresses the dissipated work as a functional of the electronic densities~\cite{Palamara2024}, or, equivalently, in terms of the static susceptibility~\cite{vn83-mt2v,PhysRevE.89.062103}.
The irreversible entropy production exhibits a pronounced peak that traces the crossover between the MI and BI regimes and provides a thermodynamic signature of the BOI region. 
This enhancement originates from the increased sensitivity of the thermal densities to variations of the control parameter $v_0$ in the vicinity of the competing MI and BI phases ($U \!\sim\! 2v_0$). 
The resulting finite-temperature phase diagram displays clear qualitative agreement with established results for the ionic Hubbard model obtained via density-matrix renormalization-group and mean-field approaches~\cite{PhysRevLett.83.2014,Tincani2009,PhysRevB.70.155115}, while exhibiting deviations in the small-$U$ and small-$v_0$ region due to thermal effects.
Next, we extend the analysis beyond the sudden-quench regime by incorporating finite-time evolution effects within linear response, approximating the relaxation function in Eq.~\eqref{eq:Psi-total} using the thALDA scheme.
Figures~\ref{fig:cumulants}(a)-\ref{fig:cumulants}(c) show the first three cumulants of the dissipated-work distribution for $U = 1 J$ as functions of the quench duration $\tau$ and for different initial amplitudes of the external potential $v_0$.
All three cumulants display a clear crossover between the nonadiabatic and adiabatic regimes.
In particular, the third cumulant vanishes at fixed $U$ for $\tau \gg \tau^\ast(v_0)$, consistent with the expectation that, in the adiabatic limit of linear-response theory, the dissipated-work distribution (generally non-Gaussian in character) approaches a Gaussian form, leaving only the first two cumulants nonzero.

The first and second cumulants, however, retain clear signatures of the transition from the BOI phase to the band-insulator phase, which persist even in the adiabatic regime.
This behavior originates from the adiabatic contribution $\beta k^{1}_{\msc{w}\,\ad}$ to the irreversible work and to the second cumulant $\beta^2 k^{2}_{\msc{w}\,\ad}$of its distribution, expressed as

\begin{align}   \label{eq:cumulants-ad}
  \beta^m k^{m}_{\msc{w}\,\ad}
  =& -m\beta^m
    \sum_{ij} \delta v_i \delta v_j
    \frac{\partial n_i^{\beta,0}}{\partial v_j} \\
 & - 2m\beta^m
    \sum_{ij} g_i g_j
    \int \frac{d\omega}{\sqrt{2\pi}}\,
      \frac{\mathrm{Im}\,\chi_{ij}(\omega)}{\omega}\,b(\omega),
\nonumber
\end{align}
for $m = 1, 2$.
At the boundary between the {BOI} region and the BI region, this term provides a substantial contribution to the total thermodynamic irreversibility.
From the behavior of all three cumulants, it can also be inferred that, in the band-insulating regime ($v_0 \gg U$), the system attains adiabaticity at faster quench rates than in either the Mott or {BOI} phases.
As a final remark, distributions satisfying the Evans-Searles fluctuation theorem~\cite{EvansSearles2002} are known to be constrained by the thermodynamic uncertainty relation~(TUR), which, in the linear-response regime, imposes a lower bound on the Fano factor,
\begin{equation}
  F_W = \frac{\mathrm{Var}(W)}{\langle W_{\diss}\rangle}
  \ge \frac{2}{\beta}.
  \label{eq:TUR}
\end{equation}
In contrast to the classical case, and in agreement with previous analyses~\cite{Guarnieri2024}, the TUR cannot be saturated in the presence of quantum fluctuations.
To examine this behavior, we computed the Fano factor for the many-body system under study.
Figure~\ref{fig:cumulants}(d) shows its dependence on the quench duration for different initial amplitudes of the staggered potential.
In the adiabatic limit (for quench times longer than the system relaxation time and at fixed temperature), $F_W$ is indeed bounded from below by $2/\beta$, as expected.
In all other regimes, however, and consistently with the generalized linear-response theory for the full quantum work statistics, this bound remains unsaturated due to quantum fluctuations.

\section{Conclusions and Discussion }

The LR-thTDDFT strategy presented here establishes a fully microscopic and predictive framework for computing the cumulants of the dissipated-work distribution in interacting many-body systems driven by spatially inhomogeneous, time-dependent external potentials. By expressing the relaxation function directly in terms of density-density response functions, this approach provides first-principles access to the full quantum work statistics within the linear-response regime, thereby avoiding phenomenological modeling strategies. A defining feature of the framework is the representation of the relaxation function through an exchange-correlation kernel, which enables the systematic incorporation of both many-body interactions and the spatial structure of the driving protocol. This formulation yields a direct connection between nonequilibrium thermodynamic irreversibility and the microscopic response properties of correlated quantum systems. 

The generality and effectiveness of LR-thTDDFT are illustrated for the one-dimensional Hubbard model in the presence of a staggered external field, where the moments of the dissipated-work distribution are evaluated from the sudden-quench regime to the adiabatic limit. Within the thALDA, distinct thermodynamic signatures of the Mott-insulating, bond-order insulating, and band-insulating phases are captured in the behavior of the lowest-order cumulants.
For paradigmatic strongly correlated systems such as those described by the Fermi-Hubbard model, a variety of complementary many-body techniques, including density matrix renormalization group, dynamical mean-field theory, and continuous-time quantum Monte Carlo have been employed  to accurately capture correlation effects. 
While the present work does not aim at a direct comparison with these approaches, previous studies have systematically benchmarked DFT-based methods against, for example, density matrix renormalization group calculations across a broad range of regimes, highlighting both their strengths and their limitations~\cite{PhysRevB.82.014202,PhysRevB.84.035111,PhysRevB.82.245114,pauletti2024quantum}.
In particular, commonly used DFT approximations, e.g. based on local density approximations,  may become unreliable in regimes dominated by strong electron-electron interactions, where the system departs significantly from an effective noninteracting description. Considerable efforts have therefore been devoted to extending the applicability of first-principles approaches to strongly correlated systems, for instance through the development of improved exchange-correlation functionals or alternative theoretical frameworks. 
Among these, recent approaches based on reformulations of the KS scheme in terms of fractionalized degrees of freedom have been proposed as a promising route to overcome some of the intrinsic limitations of conventional approximations~\cite{PhysRevLett.134.136505}. In the present work, the adiabatic local-density approximation is adopted; while retaining the simplicity and efficiency of a local and static functional, this choice should be viewed in light of the limitations discussed above. 
Importantly, this does not represent a fundamental restriction of the LR-thTDDFT framework, which naturally allows for systematic improvements through the inclusion of more refined, possibly nonlocal and frequency-dependent exchange-correlation kernels.
In this perspective, the favorable polynomial scaling of thDFT, and, by extension, thTDDFT, constitutes a key practical advantage over more computationally demanding methods, making the present framework particularly suitable for investigating nonequilibrium thermodynamics in large and realistic systems, including ultracold atomic gases and extended low-dimensional materials.

Future work will benefit from the adoption of improved approximations for the Hxc kernel, enabling an even more detailed  more detailed investigation of the behavior of the cumulants of the dissipated-work distribution under finite-time driving protocols across quantum critical points. While such protocols are already considered in the present study, no attempt is made here to extract universal scaling laws. In particular, establishing the emergence of Kibble-Zurek scaling would require a dedicated analysis, including an accurate determination of the critical lines and a careful characterization of the associated dynamical response functions.

\section*{Acknowledgements}
We acknowledge N. Lo Gullo for valuable discussions during the development of this work.
This research was partially supported by the Centro Nazionale di Ricerca in High-Performance Computing, Big Data and Quantum Computing, PNRR 4 2 1.4, CI CN00000013, CUP H23C22000360005, and by the PNRR MUR project
PE0000023-NQSTI through the secondary projects “QuCADD” and “ThAnQ”. A.P. acknowledges support from the Erasmus+ programme.

\bibliography{Refs.bib}
\appendix
\begin{figure*}
  \centering
  \includegraphics[width=0.8\textwidth]{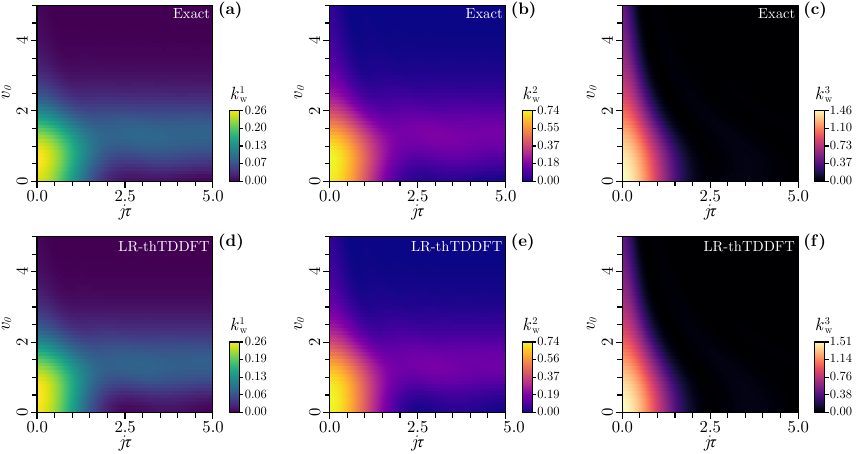}
  \caption{Benchmark of the thermal LR-TDDFT approach against exact diagonalization for the two-site Hubbard model in the canonical ensemble at half filling and zero total spin projection. Panels (a)-(c) show the first three cumulants of the dissipated-work distribution obtained from exact diagonalization, while panels (e)-(g) show the corresponding results from LR-thTDDFT. The mean relative errors are approximately $1.36\times 10^{-2}$ for the first cumulant, $2.76\times 10^{-2}$ for the second, and $1.76\times 10^{-2}$ for the third.}
  \label{fig:dimer-canonical}
%
  \includegraphics[width=0.8\textwidth]{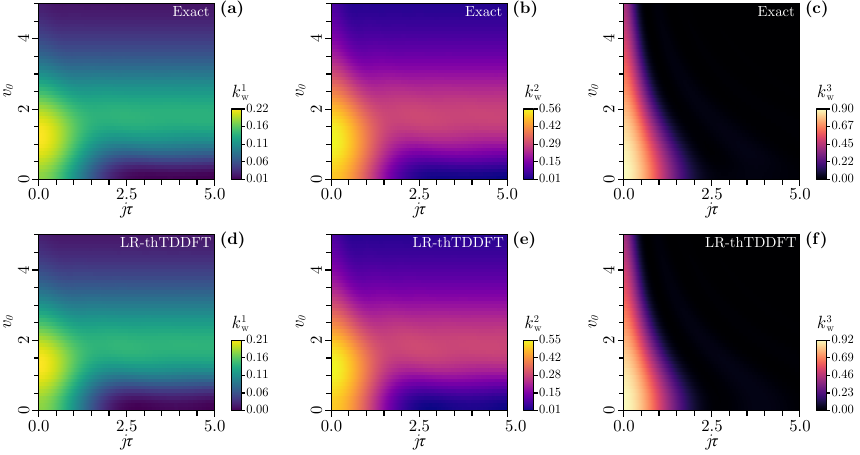}
  \caption{Benchmark of the LR-thTDDFT approach against exact diagonalization for the two-site Hubbard model in the grand-canonical ensemble. The average relative errors between the exact and LR-thTDDFT results are approximately $1.20\times 10^{-2}$ for the first cumulant, $2.45\times 10^{-2}$ for the second, and $1.61\times 10^{-2}$ for the third. Despite the larger fluctuations expected in the grand-canonical setup, the agreement remains good, confirming the robustness of the thermal linear-response TDDFT approach.}
  \label{fig:dimer-grandcanonical}
\end{figure*}

\section{Benchmark of the LR-thTDDFT approach on the Hubbard dimer\label{app:dimer}}

To assess the performance of the LR-thTDDFT approach within the ALDA approximation, we consider the exactly solvable two-site Hubbard model. Despite its simplicity, this model captures key features of the physics discussed in the main text~\cite{PhysRevLett.114.080402} and provides a stringent test for the adopted local functional due to its intrinsic spatial inhomogeneity. In finite systems, such inhomogeneities arise naturally from finite-size effects, and local density approximations are generally expected to become more accurate only in the limit of large system sizes.
As in the main text, the analysis focuses on the half-filled case with total spin projection along the $z$ axis equal to zero, subject to a time-dependent external potential. The unperturbed Hamiltonian is given by
\begin{align}
  \hat H(t)
  =& -J \sum_{i=1}^{2} \sum_{\sigma=\uparrow,\downarrow}
      \left(
        \hat c_{i,\sigma}^\dagger \hat c_{i+1,\sigma}
        + \mathrm{h.c.}
      \right) \label{eq:H-dimer}\\
   &+ U \sum_{i=1}^{2} \hat n_{i,\uparrow}\hat n_{i,\downarrow}
    + v_0\,(\hat n_1 - \hat n_2),
  \nonumber
\end{align}
which is perturbed by a time-dependent external potential of the form
\begin{equation}
  \hat V_{\msc{ext}}(t)
  = \frac{\delta v_0\,t}{\tau}\,(\hat n_1 - \hat n_2),
  \qquad \delta v_0 \ll v_0.
  \label{eq:V-dimer}
\end{equation}


The same computational protocol as in the main text is employed, adopting identical approximations for the Hxc potential in the self-consistent solution of the KS equations and for the kernel entering the response formalism. 
Calculations are performed in both the canonical and grand-canonical ensembles. As expected for a system of only two sites, small but non-negligible differences arise between the two statistical ensembles.

In both ensembles, the agreement between the exact and thermal LR-TDDFT results is very good at both qualitative and quantitative levels, as shown in Figs.~\ref{fig:dimer-canonical} and~\ref{fig:dimer-grandcanonical}. This indicates that the approach captures the essential features of the nonequilibrium thermodynamic response even in strongly correlated regimes.
Further improvements in accuracy would require going beyond the adiabatic local-density approximation by incorporating an explicit frequency dependence in the Hxc kernel. Such an extension would allow the framework to capture dynamical correlations and a broader spectrum of many-body excitations beyond the local and static limits. 
While the present treatment already enables the description of quantum phase transitions in the present case, an explicit frequency dependence in the Hxc kernel is indeed expected to be important in the vicinity of phase transitions, where enhanced fluctuations and nonlocal dynamical effects play a central role.


\end{document}